\documentclass[conference,a4paper]{APSIPA2018}
\usepackage{multirow}
\usepackage{amsmath,graphicx}
\usepackage[psamsfonts]{amssymb}
\usepackage{bm}
\usepackage{amsxtra}
\usepackage{threeparttable}
\usepackage{subfig}
\usepackage{flushend}
\graphicspath{{./pdf/}}
\DeclareGraphicsExtensions{.pdf}

\newcommand{\tabincell}[2]{\begin{tabular}{@{}#1@{}}#2\end{tabular}}

\def\vec#1{\ensuremath{\bm{{#1}}}}
\def\mat#1{\vec{#1}}

\begin{document}

\title{Investigation of Frame Alignments for GMM-based Digit-prompted Speaker Verification}

\author{%
\authorblockN{%
Yi Liu\authorrefmark{1},
Liang He\authorrefmark{1},
Weiqiang Zhang\authorrefmark{1}
Jia Liu\authorrefmark{1},
Michael T. Johnson\authorrefmark{2}
}
\authorblockA{%
\authorrefmark{1}
Tsinghua National Laboratory for Information Science and Technology, \\
Department of Electronic Engineering, Tsinghua University, Beijing 100084, China \\
E-mail: liu-yi15@mails.tsinghua.edu.cn, \{heliang, wqzhang, liuj\}@tsinghua.edu.cn}
\authorblockA{%
\authorrefmark{2}
Department of Electrical and Computer Engineering, University of Kentucky, \\
453 F. Paul Anderson Tower, Lexington, KY 40506-0046, USA \\
E-mail: mike.johnson@uky.edu}
}

\maketitle
\thispagestyle{empty}

\begin{abstract}
Frame alignments can be computed by different methods in GMM-based speaker verification. By incorporating a phonetic Gaussian mixture model (PGMM), we are able to compare the performance using alignments extracted from the deep neural networks (DNN) and the conventional hidden Markov model (HMM) in digit-prompted speaker verification.
Based on the different characteristics of these two alignments, we present a novel content verification method to improve the system security without much computational overhead. Our experiments on the RSR2015 Part-3 digit-prompted task show that, the DNN-based alignment performs on par with the HMM alignment. The results also demonstrate the effectiveness of the proposed Kullback-Leibler (KL) divergence based scoring to reject speech with incorrect pass-phrases.
\end{abstract}

\section{Introduction}
Automatic speaker verification (ASV) has developed rapidly in the last decade. ASV can be broadly categorized into text-independent and text-dependent applications. In text-independent ASV, the system verifies the speaker's identity using spontaneous speech, which is valuable for military and forensic tasks. Text-dependent ASV, on the other hand, requires users to utter a specific pass-phrase, and is commonly used in commercial applications.

Although text-independent speaker verification is much more flexible, text-dependent ASV is more suitable for applications which require higher security. The pass-phrases in text-dependent speaker verification can be fixed or prompted during enrollment and test, with text-prompted the more popular approach in recent years \cite{rsr2015}.
Text-prompted systems requires the user to both speak the correct text and be validated as the target speaker. For instance, in digit-prompted ASV, users are required to speak different digit strings. By verifying the text contents, text-prompted speaker verification provides users an additional protection and is more robust to replay spoofing attacks \cite{antispoof}. In this paper, we focus on the digit-prompted case and investigate both the speaker and content verification.

By virtue of a constrained vocabulary, digit-prompted speaker verification usually delivers better performance for short utterances. Many modeling methods have been developed for digit-prompted speaker verification. The most commonly used approach is based on Gaussian mixture models (GMMs), motivated by text-independent speaker verification.
Universal background model (UBM) based maximum a posteriori (MAP) \cite{gmm_hmm}, joint factor analysis \cite{hmm_jfa}, i-vector with within class covariance normalization (WCCN) \cite{hmm_ivec} and probabilistic linear discriminant analysis (PLDA) \cite{plda} have all been applied to this field.

The calculation of the frame posterior (often referred as \emph{frame alignment}) plays an essential role in these GMM-based models. The frame alignment $P(s|\vec{x}_t)$ is the posterior probability that speech frame $\vec{x}_t$ belongs to a phonetic unit $s$. Depending on the alignment methods, the phonetic units can either be the GMM components generated by unsupervised clustering or be assigned to the classes used in speech recognition (e.g., monophones, monophone states or senones). With more accurate frame alignment, the speaker verification system can better model the phonetic units, and effectively compare the features belonging to the same feature subspace. In text-independent speaker verification, researchers have used senones predicted by phonetically-aware deep neural networks (DNNs) to generate frame alignments and improved the performance significantly \cite{lei_dnn}.

In digit-prompted speaker verification, if we assume all users speak the correct digit strings, the transcriptions of the utterances are known beforehand. As in speech recognition, Viterbi forced alignment based on hidden Markov models (HMMs) is a natural choice to obtain the alignments \cite{gmm_hmm}. Using the speech content, the HMM alignment provides accurate results even in adverse acoustic environments. However, the user may incidentally misread the prompted text. This mismatch will severely impact the quality of the HMM alignment.


As an alternative to the HMM alignment, conventional DNN alignment can also be applied to extract the posteriors in digit-prompted speaker verification. It has been found that i-vector modeling using DNN alignment achieves good performance on RSR2015 Part-3 evaluation \cite{dnn_rsr2015}. In \cite{hmm_dnn_odyssey, hmm_dnn_ivec}, the authors showed that DNN alignments outperformed HMMs on both RSR2015 and RedDots \cite{reddots}. Note that the quality of the DNN alignment is influenced by the acoustic environment rather than the transcriptions.

In this paper, we first compare the performance of HMM and DNN alignments, using both GMM-MAP and i-vector modeling, on the RSR2015 Part-3 digit-prompted task. To better understand the difference between these two alignments, the HMM and DNN are trained on the same dataset and based on the same phonetic units. Due to the limited number of phonetic units in the digit-prompted verification, we use a phonetic GMM (PGMM) approach \cite{ti_short} for both GMM-MAP and i-vector modeling, keeping roughly the same number of Gaussian mixtures across the different systems.

The assumption that all the input utterances contain the correct contents is unrealistic. Attackers can record the speech of a target user, and malicious replay attacks often present the speech with wrong text content. In fixed-phrase text-dependent ASV, techniques like keyword spotting or wakeup word detection \cite{chen2015query} can be applied to verify the content. While on the digit-prompted condition, speech recognition is the most commonly used method, which introduces extra complexity in the computation and the system deployment. Since the HMM and DNN alignments exhibit complementary properties, the sequence information in both alignments can be considered together. In this paper, we propose a novel Kullback-Leibler (KL) divergence based scoring  to verify the text content without speech recognition decoding. This algorithm involves low computational overhead and is especially suitable for the embedded devices. The effectiveness of this method is validated by our experiments.  

The organization of this paper is as follows. Modeling methods based on the HMM alignment are briefly introduced in Section 2. Section 3 describes PGMM using the DNN alignment. Then, based on these alignment approaches, a fast and efficient algorithm to verify the utterance content is proposed in Section 4. Experimental setup and results are presented in Section 5 and 6. Finally, Section 7 concludes the paper.

\section{The Role of HMM}
In GMM-based speaker verification, a frame alignment is first generated and then used to calculate the Baum-Welch statistics. In digit-prompted speaker verification, the Viterbi or forward-backward (FB) algorithm is a common choice to align speech frames to the HMM phonetic units. On the digit-prompted condition, each digit is treated as a whole-word model. In this paper, we use $N$-state HMMs to represent ten digits plus silence.

Given a transcription, a directed graph of HMM states is first compiled. The Viterbi or FB algorithm uses this graph to find a state path that optimally fits the feature sequence. In Viterbi forced alignment, frame $\vec{x}_t$ is aligned to the most likely state $q_t$, i.e. $P(s|\vec{x}_t) = 1$ if and only if $s=q_t$. The FB alignment, which can be seen as a \emph{soft} version of the Viterbi algorithm, computes the posterior from forward and backward probabilities \cite{hmm}. In \cite{liu_comparison}, it was shown that these two types of alignments result in similar performance. To better compare the alignments generated by HMM and DNN, the soft FB alignment is used in our experiments.

In GMM-HMM, the state $s$ is modeled by GMM $\lambda_s=\{w_{s,c},\vec{\mu}_{s,c},\mat{\Sigma}_{s,c}\}$, which is treated as the universal background model (UBM), as in the text-independent speaker verification. Frames are aligned to different Gaussian mixtures of states using HMM alignment $P(s|\vec{x}_t,h)$
\begin{equation} \label{eq:posterior}
\gamma^{\text{HMM}}_{s,c,t} = P(s|\vec{x}_t,h) P(c|\vec{x}_t,\lambda_s)
\end{equation}
where $\gamma^{\text{HMM}}_{s,c,t}$ is the posterior of $\vec{x}_t$ occupying the $c$-th mixture of state $s$, $h$ represents the HMM, $P(c|\vec{x}_t,\lambda_s)$ is the Gaussian posterior of component $c$ in GMM $\lambda_s$. 

The normalized Baum-Welch statistics using the GMM-HMM alignment are
\begin{equation} \label{eq:zero_order}
N_{s,c} = \sum_t \gamma^{\text{HMM}}_{s,c,t}
\end{equation}
\begin{equation} \label{eq:first_order}
\vec{\bar{F}}_{s,c} = \sum_t \gamma^{\text{HMM}}_{s,c,t} (\vec{x}_t - \vec{\mu}_{s,c})
\end{equation}
\begin{equation} \label{eq:second_order}
\mat{\bar{\Sigma}}_{s,c} = \sum_t \gamma^{\text{HMM}}_{s,c,t} (\vec{x}_t - \vec{\mu}_{s,c}) (\vec{x}_t - \vec{\mu}_{s,c})^T
\end{equation}
To adapt the HMM to a speaker model, GMM-MAP estimation is applied as
\begin{equation} \label{eq:map}
\vec{\hat{\mu}}_{s,c} = \alpha_{s,c} \vec{\bar{F}}_{s,c} + \vec{\mu}_{s,c}
\end{equation}
where $\vec{\hat{\mu}}_{s,c}$ is the mean of the adapted speaker model and $\alpha_{s,c} = 1/(N_{s,c}+r)$, $r$ is the relevance factor \cite{gmm_map}. During test, the verification score is the log-likelihood ratio computed against the speaker model and UBM.

In the i-vector modeling, the statistics extracted by (\ref{eq:zero_order}-\ref{eq:second_order}) are used for the total variability space training and i-vector extraction \cite{ivector}. The flowchart of GMM-MAP and i-vector using the GMM-HMM alignment is illustrated in Fig. \ref{fig:system_diagram}a. In the figure, speaker and speech features indicate the features more suitable for speaker and speech recognition, respectively.

In DNN-HMM, each state is modeled by a DNN output. The state posterior $P(s|\vec{x}_t,h)$ is calculated based on the log-likelihoods computed by a DNN rather than several GMMs. In this case, the within-state posterior  $P(c|\vec{x}_t,\lambda_s)$ cannot be calculated directly. It should be estimated using the method introduced in the next section.

\section{PGMM with DNN-based Alignment}
In conventional text-independent speaker verification, the outputs of the DNN are usually senones modeled by a single Gaussian \cite{lei_dnn}. However, in digit-prompted applications, the number of states is much smaller than for the text-independent case due to the limited vocabulary \cite{gmm_hmm, dnn_rsr2015}. To increase the modeling capability of the DNN alignment, a DNN-based phonetic GMM (PGMM) is used in this paper. Each state in the PGMM is represented by a GMM rather than a single Gaussian. The idea of PGMM was first proposed in \cite{asr_pgmm}, and was once applied to DNN/i-vector framework in \cite{dnn_pgmm}.

\begin{figure}[tb]
\begin{minipage}[b]{0.9\linewidth}
  \centering
  \centerline{\includegraphics[width=0.85\linewidth]{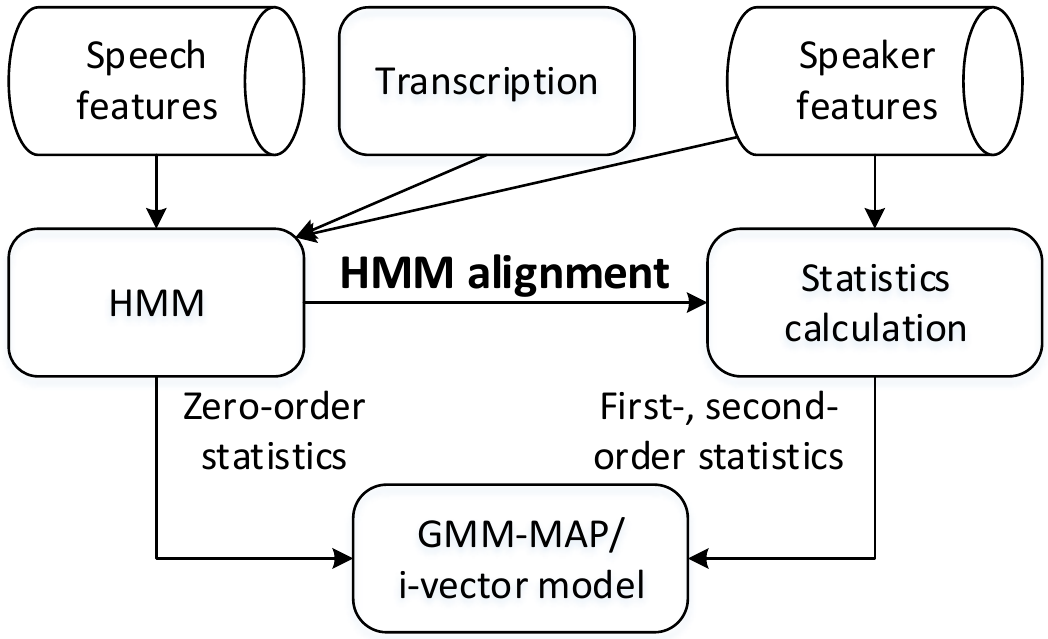}}
  \centerline{(a)}\medskip
  \label{fig:test}
\end{minipage}
\begin{minipage}[b]{0.9\linewidth}
  \centering
  \centerline{\includegraphics[width=0.85\linewidth]{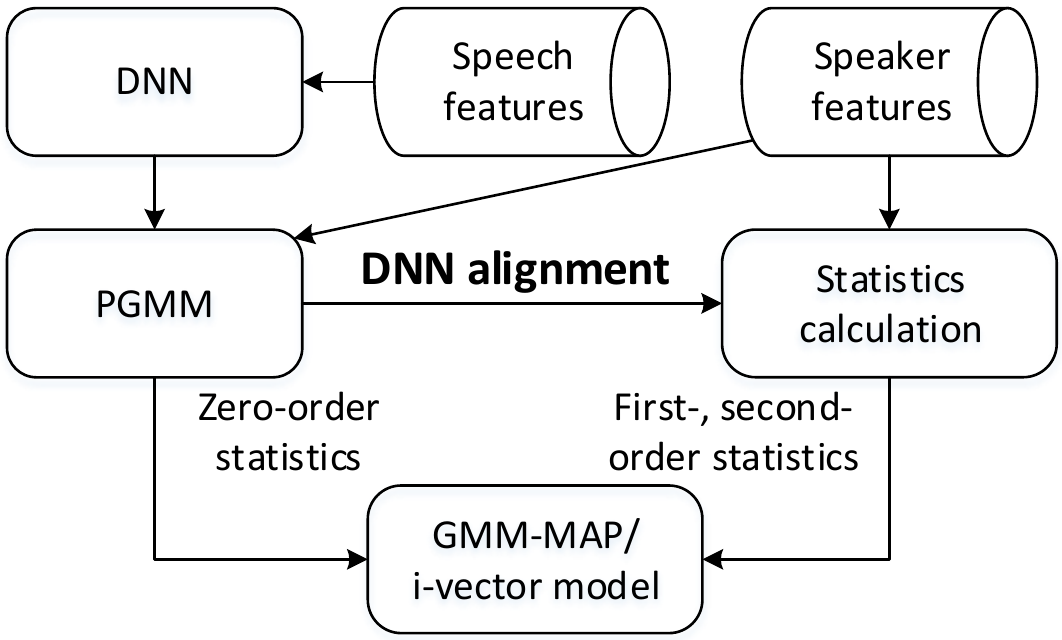}}
  \centerline{(b)}\medskip
\end{minipage}
\caption{The flow diagrams of digit-prompted speaker verification systems based on (a) HMM alignment and (b) DNN alignment.}
\label{fig:system_diagram}
\end{figure}

To initialize the model, each feature is first hard aligned to one states with the maximum posteriors. The initial state GMMs are then trained individually.
Let $\tau$ denote the DNN model, and $\gamma^{\text{DNN}}_{s,c,t}$ is the posterior of $\vec{x}_t$ occupying the $c$-th mixture of DNN state $s$. Using the DNN alignment  these GMMs can be trained by EM algorithm and the sufficient statistics are accumulated by (\ref{eq:pgmm_e_step}-\ref{eq:pgmm_m_step_cov}).
\begin{equation} \label{eq:pgmm_e_step}
\gamma^{\text{DNN}}_{s,c,t} = P(s|\vec{x}_t,\tau) P(c|\vec{x}_t, \lambda_s)
\end{equation}
\begin{equation} \label{eq:pgmm_m_step_zero_order}
N_{s,c} = \sum_t \gamma^{\text{DNN}}_{s,c,t}
\end{equation}
\begin{equation} \label{eq:pgmm_m_step_weight}
w_{s,c} = N_{s,c} / \sum_{c} N_{s,c}
\end{equation}
\begin{equation} \label{eq:pgmm_m_step_mean}
\vec{\mu}_{s,c} = \frac{1}{N_{s,c}} \sum_t \gamma^{\text{DNN}}_{s,c,t} \vec{x}_t
\end{equation}
\begin{equation} \label{eq:pgmm_m_step_cov}
\mat{\Sigma}_{s,c} = \frac{1}{N_{s,c}} \sum_t \gamma^{\text{DNN}}_{s,c,t} (\vec{x}_t - \vec{\mu}_{s,c})(\vec{x}_t - \vec{\mu}_{s,c})^T
\end{equation}
The parameters of PGMM are updated iteratively. The PGMM algorithm can also be used in DNN-HMM to calculate the within state posterior, by replacing the DNN posteriors $P(s|\vec{x}_t,\tau)$ with the alignment posteriors $P(s|\vec{x}_t,h)$. 

As with the HMM alignment, the posteriors $\gamma^{\text{DNN}}_{s,c,t}$ calculated from (\ref{eq:pgmm_e_step}) are used to extract the statistics. GMM-MAP and i-vector modeling are applied in the same manner as described in Section 2, expect for the different alignment sources. The procedure is demonstrated in Fig. \ref{fig:system_diagram}b.

\section{Fast Content Matching}
In real-world applications, speaker verification systems face many kinds of attacks \cite{antispoof}. Replay is a form of low-cost spoofing attack where an adversary claims to be a target speaker using a recorded speech sample. If the text is prompted in the runtime, we can prevent this replay attack effectively, since these samples are often captured surreptitiously and the attacker cannot carefully control the content. Hence the content matching is important to enhance the security in commercial applications.

However, most published papers about text-prompted speaker verification do not focus on content verification, often treating this as a separated speech recognition task. When pass-phrases are fixed during enrollment and test, some techniques, e.g., query-by-example (QbE) keyword spotting \cite{chen2015query} or dynamic time wrapping (DTW) \cite{post_dtw,ivec_dtw}, can be used to compare the contents. However, these techniques cannot be easily applied to the digit-prompted case considering that the pass-phrases in the enrollment and test differ from each other. Another option to do the content verification is to insert a speech recognition system \cite{gmm_hmm}. The disadvantage of this method is that the decoding procedure costs significant time and increases the model complexity.

In the PGMM framework, the DNN is trained based on the HMM alignment. Although they share the same set of phonetic units, they generate the alignments from different perspectives.
The HMM is capable to find the globally optimal alignment given the transcription. If the transcription is consistent with the speech content (which means the user speaks the correct prompt), the HMM generates the alignment with good quality. If the content is inconsistent, the alignment becomes incorrect respectively. In contrast, the DNN aligns the frames only based on the local acoustic features and will not be impacted by any transcriptions. Considering these different characteristics of the two alignments, we use the DNN alignment as a reference and propose a fast and efficient scoring method to verify the text content without any decoder.

The alignments are posteriors from the HMM and the DNN. The deviation between these posteriors can be defined by Kullback-Leibler (KL) divergence:
\begin{equation} \label{eq:kld}
\text{KL} = \frac{1}{T} \sum^{T}_{t=1} \sum_p \gamma^{\text{HMM}}_{p,t} \log \left( \gamma^{\text{HMM}}_{p,t} / \gamma^{\text{DNN}}_{p,t} \right)
\end{equation}
where $T$ is the total number of frames and $p$ denotes the phonetic class which will be explained later.

The KL divergence in (\ref{eq:kld}) should be small when the speech is correctly uttered, since the HMM and DNN alignments both exhibit the genuine phonetic sequence.
Biases will be introduced into the HMM alignment when the transcription is uttered incorrectly, making the alignment sequence deviate from that of the DNN. The mismatch leads to a increase in the KL divergence. Thus, the deviation between these two posteriors becomes a metric to detect speech with wrong text.

The straightforward definition of a phonetic class $p$ in digit-prompted speaker verification is a single state $s$ of a digit, but we find it is always beneficial to involve a broader class, i.e., $p=\{s|s \in d_p\}$, $d_p$ is a digit. The rational is easy to understand. Since the DNN alignment is noisy and can be harassed by the adverse acoustic environment, the broader definition improves the robustness to some alignment errors.

Given the definition, the posterior of a phonetic class $p$ at time $t$ is expressed as:
\begin{equation} \label{eq:phonetic_unit}
\gamma^{\{\cdot\}}_{p,t} = \sum_{s\in p}\sum_c \gamma^{\{\cdot\}}_{s,c,t}
\end{equation}
where $\{\cdot\}$ is HMM or DNN respectively.

Note that the KL divergence is defined only if $\gamma^{\text{DNN}}_{p,t} = 0$ implies $\gamma^{\text{HMM}}_{p,t}=0$. The HMM and DNN alignments are sparse, and may violate this condition. To alleviate this problem, we adjust the both posteriors by
\begin{equation} \label{eq:post_adjust}
\gamma'_{p,t} = \frac{\gamma_{p,t} + \epsilon}{\sum_p  (\gamma_{p,t}+\epsilon)}
\end{equation}
where $\epsilon=10^{-5}$ is a small constant.

By this method, the content verification becomes simply a byproduct of the speaker verification. An additional HMM alignment is the only extra step to generate a content verification score, after the DNN posteriors computation. Alignment algorithm based on Viterbi or FB is much faster and easier to implement than a speech recognition decoder \cite{htk_book}. Unlike QbE or DTW, which only works for the fixed-phrase condition, KL divergence scoring can be applied to content verification with arbitrary text. This light-weight algorithm is suitable for embedded devices whose power and computing capability is limited.

\section{Experimental Setup}
\subsection{Data}
The experiments are carried out on RSR2015 Part-3, which is a digit-prompted task. Digit strings are prompted during enrollment and test. The enrollment data for a speaker contains three ten-digit utterances, and the test utterance is a five-digit sequence. The background and development sets (about 22 hours) are used for the gender-independent UBM, DNN, PGMM and i-vector training. The evaluation set contains 57 male and 49 female speakers. The detailed statistics of this dataset can be found in \cite{rsr2015}. The trials can be partitioned into four categories: (1) target speaker produces correct content (TC), (2) target speaker produces wrong content (TW), (3) imposter speaker produces correct content (IC), and (4) imposter speaker produces wrong content (IW). Only the first TC combination is accepted by the digit-prompted system, while the other three categories are all non-target trials. Equal error rate (EER) and minimum decision cost function (minDCF) in SRE08 \cite{sre08} and SRE10 \cite{sre10} are used for evaluation. For speaker verification, we investigate the performance of TC-IC (i.e. compute the metrics among TC and IC trials), and the performance of TC-TW is taken into account when we check the effectiveness of the proposed fast content verification method. We do not report TC-IW, because it is the easiest task that all systems achieve relatively good performance.

\subsection{Models}

\begin{itemize}
   \item \textbf{DNN}: The DNN in this paper consists of 4 fully connected layers with 512 nodes per layer. The input is a vector of 120-dimensional FBank (40 filter-bank energies plus delta/delta-delta) features with symmetric 5-frame expansion, resulting in 11 frames in total. The number of output nodes is 33, which equals to the state number in our HMM. The training data is generated by the HMM-based Viterbi alignments. This DNN is also used in the following DNN-HMM model.
  \item \textbf{HMM}: The GMM-HMM and DNN-HMM used in our experiment is trained using HTK \cite{htk_book}. Each word (ten digits plus silence) is modeled by a 3-state HMM. In GMM-HMM, each state is represented by a 16-component GMM. 60-dimensional MFCC features (20 static + delta/delta-delta) with CMVN are used to train the GMM-HMM. The MFCC features are also used in all models other than the DNN. 
  \item \textbf{Conventional GMM-MAP/i-vector}: A gender-independent UBM with 512 mixtures is trained. The relevance factor is 5.0 in the GMM-MAP system. For i-vector modeling, the rank of the i-vector subspace matrix is 400. LDA, length normalization and PLDA are applied to score the trials.
  \item \textbf{PGMM}: For DNN and DNN-HMM alignments, the PGMM model is used to calculate the statistics. States corresponding to silence are ignored. Each state is represented by a 16-component GMM. The total number of mixtures in the PGMM is 480. All other configurations are the same as the conventional GMM-MAP/i-vector.
\end{itemize}

\section{Results}
\subsection{Comparison of different alignments}

The speaker verification performance of GMM-MAP and i-vector on TC-IC trials are presented at the top and bottom parts of Table \ref{table:result_alignment}. Within all the alignments, we find that GMM-MAP consistently outperforms i-vector. Since i-vector is only the point estimation of the latent factor in the total variability subspace, it becomes less reliable due to the large variance when the duration is too short and it cannot achieve good performance in this task. Also as expected, frame alignments calculated on the unsupervised GMM perform the worst in both modeling methods.

\begin{table}[htb]
\caption{Performance (EER(\%)/minDCF08/minDCF10) of GMM-MAP and i-vector using different frame alignments. Only TC-IC is reported.}
\label{table:result_alignment}
\centerline{
\resizebox{\linewidth}{!}{\begin{tabular}{|c|c|c|}
\hline
\multirow{2}{*}{Models} & male & female \\
\cline{2-3}
 & TC-IC & TC-IC \\
\hline
\hline
GMM-MAP          & 3.38 / 0.0165 / 0.6607 & 3.36 / 0.0161 / 0.4951 \\
\hline
DNN/GMM-MAP      & 2.08 / 0.0115 / 0.5307 & \textbf{2.68 / 0.0129 / 0.4525} \\
\hline
HMM/GMM-MAP      & 2.20 / 0.0116 / 0.5128 & 3.16 / 0.0140 / 0.4676 \\
\hline
DNN-HMM/GMM-MAP  & \textbf{2.06 / 0.0112 / 0.5173} & 2.95 / 0.0137 / 0.4565 \\
\hline
\hline
i-vector         & 3.49 / 0.0174 / 0.5686 & 3.50 / 0.0176 / 0.5467 \\
\hline
DNN/i-vector     & \textbf{2.74 / 0.0148 / 0.5357} & \textbf{3.10 / 0.0169 / 0.5740} \\
\hline
HMM/i-vector     & 2.77 / 0.0162 / 0.6583 & 3.10 / 0.0170 / 0.6154 \\
\hline
DNN-HMM/i-vector & 2.71 / 0.0162 / 0.6045 & 3.34 / 0.0169 / 0.5988 \\
\hline
\end{tabular}
}
}
\end{table}

In Table \ref{table:result_alignment}, it is interesting to find that the DNN-based alignment achieves results on par with the HMM-based alignments. Actually, it outperforms GMM-HMM on many conditions. Even though DNN-HMM can achieve a better alignment quality than GMM-HMM theoretically, it fails to beat the DNN alignment as well. The results are consistent with \cite{hmm_dnn_ivec}. Unlike \cite{hmm_dnn_ivec}, we use the same dataset to train the HMM and DNN and the basic phonetic units are also the same. The comparison should be fair in this case.

This experiment indicates that the extra text information adds little to speaker verification results on RSR2015 Part-3. We hypothesize the reason is that, the local information used in the DNN aligns the frames to the phonetic units well, while the HMM alignment are likely to be disturbed. For example, users may mispronounce the phones in the training data which leads to poor HMM alignments.

\subsection{Reject wrong text}
Although the HMM alignment does not boost the performance of speaker verification, the prompted pass-phrases can still help to improve the security. The KL divergence in (\ref{eq:kld}) is used to reject non-matched content. GMM-HMM and DNN-HMM are two sources of HMM alignments while the DNN provides the reference. As a comparison, a speech recognition decoder is implemented by HTK and the method in \cite{gmm_hmm} is used as the baseline. For speech recognition decoder, the acoustic model is the GMM-HMM used in the last experiment, and the language model is a word-loop constructed by ten digits plus silence. This decoding-based text verification is also shown in \cite{gmm_hmm}.

As described in Section 4, two different phonetic classes are investigated. The \emph{state-level} in Table \ref{table:result_kld} denotes that the KL divergence is computed between states, while the states belonging to the same digit are clustered as one phonetic class at the \emph{digit-level} systems. Results on the TC-TW trials are shown in Table \ref{table:result_kld}.

\begin{table}[htb]
\caption{Performance (EER(\%)/minDCF08/minDCF10) of different content verification methods. Results are reported on TC-TW trial.}
\label{table:result_kld}
\centerline{
\resizebox{\linewidth}{!}{\begin{tabular}{|c|c|c|}
\hline
\multirow{2}{*}{Methods} & male & female \\
\cline{2-3}
 & TC-TW & TC-TW \\
\hline
\hline
\tabincell{c}{Decoding \cite{gmm_hmm} \\ }       & 0.25 / 0.0010 / 0.0704 & 0.05 / 0.0004 / 0.0394 \\
\hline
\hline
\tabincell{c}{DNN+GMM-HMM (state)}  & 0.74 / 0.0041 / 0.1721 & 0.37 / 0.0027 / 0.1303 \\
\hline
\tabincell{c}{DNN+DNN-HMM (state)}  & 0.40 / 0.0016 / 0.0699 & 0.03 / 0.0003 / 0.0312 \\
\hline
\tabincell{c}{DNN+GMM-HMM (digit)}  & 0.30 / 0.0016 / 0.0566 & 0.03 / 0.0002 / 0.0250 \\
\hline
\tabincell{c}{DNN+DNN-HMM (digit)}  & \textbf{0.25 / 0.0012 / 0.0587} & \textbf{0.02 / 0.0002 / 0.0210} \\
\hline
\end{tabular}
}
}
\end{table}

As shown in Table \ref{table:result_kld}, the decoding-based method performs well in this task. Meanwhile, the proposed KL divergence based scoring is also an efficient way to verify the text content. Compared to the decoding-based method, the KL divergence achieves competitive performance on both male and female trials. Among all the results, scoring between DNN and DNN-HMM performs better. The reason is that when aligning frames using DNN-HMM, we use the posteriors generated by the DNN, making their posteriors more consistent with each other. Thus the deviation for matched speech can be minimized.

Different phonetic classes are also investigated. In practice, we observed that the DNN alignment can differ from the HMM alignment due to the adverse acoustic environment and the mispronunciation. A broader phonetic class is more robust to this disturbance. It is presented in Table \ref{table:result_kld} that, compared to the state-level class, using digits as the phonetic classes illustrates better performance. Overall, the digit-level KL divergence scoring using DNN and DNN-HMM alignments achieves the best performance.

\section{Conclusions}
In this paper, we first compare several speaker verification methods implemented on the RSR2015 Part-3 digit-prompted task. The experiments show that GMM-MAP outperforms vanilla i-vector on this task. In both modeling methods, posteriors calculated on DNN and HMM achieve better performance than unsupervised GMM. Alignments based on DNN generally exhibit better performance than GMM-HMM alignments. The use of DNN-HMM, which is a common practice in speech recognition, does not lead to further improvements. The results demonstrate the effectiveness of DNN alignment in digit-prompted speaker verification. In contrast, the use of HMM seems to be affected by inaccurate pronunciation and mismatched text and do not outperform the DNN alignment.

Considering the characteristics of these different alignments, we then propose a novel algorithm to further verify the text content. The KL divergence between the DNN- and HMM-aligned posteriors efficiently conveys the content matching result, especially when DNN and DNN-HMM is used. The digit-level KL divergence scoring with broader phonetic classes is applied to further improve the robustness. This method is useful under both fixed- and prompted-phrase conditions. Although the speech recognition decoder can also be used, we note that this KL divergence scoring is much faster and easier to implement.

In the future, we will explore new methods to fully utilize the known transcriptions in the text-prompted task. Also, we will develop the KL divergence based method to verify trials with a larger vocabulary.

\section*{Acknowledgment}
The work is supported by National Natural Science Foundation of China under Grant No. 61370034, No. 61403224 and No. 61273268.

\bibliographystyle{IEEEbib}
\bibliography{myrefs}

\end{document}